\documentclass[%
 reprint,
 amsmath,amssymb,
 aps,
]{revtex4-2}
\usepackage{draftwatermark}
\SetWatermarkText{}
\SetWatermarkScale{5}
\usepackage{graphicx}
\usepackage{dcolumn}
\usepackage{bm}

\usepackage{xcolor}
 \usepackage{multirow}
\begin{document}

\preprint{APS/123-QED}

\title{Thermal Gradient-Driven Skyrmion Dynamics with Near-Zero Skyrmion Hall Angle}

\author{Yogesh Kumar}
 \altaffiliation[Corresponding author:]{yogeshmalya111@gmail.com}
 \affiliation{%
Department of Physics, Indian Institute of Technology Delhi, 110016, India
}%
\author{Hurmal Saren}%
 \altaffiliation[Corresponding author:]{hurmalsaren@gmail.com}
\affiliation{%
Department of Physics, Indian Institute of Technology Delhi, 110016, India
}%
\author{Pintu Das}%
 \email{pintu@physics.iitd.ac.in}
\affiliation{%
Department of Physics, Indian Institute of Technology Delhi, 110016, India
}

\date{\today}

\begin{abstract}
Thermal-gradient-driven skyrmion dynamics offers a promising route toward green spintronics, enabling the utilization of waste heat for information transport and processing. Using micromagnetic simulations, we investigate Néel skyrmions in a Co/Pt bilayer nanoracetrack and demonstrate that stochastic torques induced by a thermal gradient drive skyrmion motion toward the hotter region with a nearly vanishing Hall angle. The dynamics depends sensitively on intrinsic material parameters: the skyrmion velocity decreases with increasing damping constant ($\alpha$), increases with stronger thermal gradients, and varies systematically with saturation magnetization, interfacial DMI strength, and uniaxial out-of plane anisotropy. Importantly, we identify a specific range of material parameters within which the skyrmion velocity changes sharply while the Hall angle remains strongly suppressed, saturating near zero. This comprehensive parameter-dependent study establishes a universal design framework for minimizing the Hall effect in thermal-gradient–driven spintronic systems.
\end{abstract}

\maketitle


\section{\label{sec:level1}Introduction}
Magnetic skyrmions are topologically protected vortex-like spin configurations, in which the spins wrap around the entire surface of the sphere exactly once. Skyrmions are stable against external perturbations due to their topological protection, and their controlled dynamics is explored for high-density magnetic storage, logic devices, neuromorphic computing, skyrmion transistors, reconfigurable magnonic devices, microwave devices, quantum information,  and energy-efficient interconnects~\cite{fert2013skyrmions, romming2013writing, liu2015dynamical, zhang2015magnetic, li2017magnetic, pribiag2007magnetic, Mühlbauer2009, sampaio13, Nagaosa13}. Skyrmions are predicted in several systems, including B20 crystals, ultrathin ferromagnets (2D heterostructures), and heavy-metal/magnetic heterostructures \cite{fert2013skyrmions, sampaio13, Back_2020, Mathur_2019, Zhang_23}. The stabilization of skyrmions occurs due to contributions from various energy terms, such as dipolar, perpendicular magnetic anisotropy, exchange interaction, and Dzyaloshinskii-Moriya interaction (DMI) \cite{sampaio13, fert2013skyrmions}.

Skyrmions can be driven by spin-polarized currents via spin-transfer torque (STT)~\cite{fert2013skyrmions, iwasaki2013, iwasaki} or spin-orbit torque (SOT)~\cite{Woo2017, Legrand2017, Peng_nat_comm_2021} mechanisms. Under such torques, skyrmions exhibit transverse motion at a certain angle due to the Magnus force, resulting in a trajectory that deviates from the direction of the driving current. This deflection, known as the skyrmion Hall effect (SHE)~\cite{Jiang16, chen17}. While controlled transverse motion can be advantageous for certain spintronic applications such as adder-subtractor circuits\cite{Yang21}, channeling in nanodevices\cite{Yang24}, Sorting \& Multiplexing\cite{Yang24}, Microwave Oscillators \cite{Yang24}, etc. SHE presents challenges for devices requiring purely longitudinal skyrmion transport~\cite{Luo219}. Mitigating SHE remains challenging, but several complex strategies have been proposed, including enhancing perpendicular magnetic anisotropy (PMA) at racetrack boundaries~\cite {zhang16}, exploiting shape anisotropy~\cite{lai17}, employing antiferromagnetically coupled skyrmions~\cite{Legrand20}, and using both PMA gradient and spintronics~\cite{de_Assis2023}.

Skyrmion dynamics have been explored using various alternative driving mechanisms, including anisotropy gradients~\cite{de_Assis2023}, thermal gradients ($\boldsymbol{\nabla} T{\rm{(x)}}$), and magnetic field gradients~\cite{Zhang}. Among these, thermally induced skyrmion dynamics~\cite{kong2013} has gained significant interest due to its relevance in understanding Joule heating and thermal fluctuations~\cite{Zhang2018}.

Recent theoretical studies have predicted a range of emergent mechanisms induced by thermal spin currents, including the spin Nernst effect~\cite{Sheng17}, spin Seebeck effect~\cite{Uchida2008}, and thermal Hall effect~\cite{Park20}. These effects generate spin currents that impose effective torques that can interact with magnetic spin texture, offering a new direction for low-dissipation control of domain walls and skyrmions. Among these, unidirectional skyrmion motion under a $\boldsymbol{\nabla} T{\rm{(x)}}$ has recently attracted significant attention due to its promise for energy-efficient spintronic devices that harness ambient or waste heat. Several theoretical and experimental studies have demonstrated $\boldsymbol{\nabla} T{\rm{(x)}}$ induced unidirectional motion of skyrmions~\cite{kong2013, Yu21, qin2022, Zhang2018}. However, whether skyrmion motion strictly follows the direction of $\boldsymbol{\nabla} T{\rm{(x)}}$ remains unresolved. Moreover, the existence of a thermal skyrmion Hall effect presents an intriguing open question that needs further exploration.

The underlying skyrmion dynamics can be understood through both microscopic and macroscopic thermodynamic theories. The effective torque driving the skyrmion results from competition among different torque, including thermal magnonic spin torques~\cite{kong2013, raj2023, qin2022}, entropic forces~\cite{Yu21, raj2023}, Seebeck-effect-induced thermal torques~\cite{wang2020}, and thermally induced dipolar fields~\cite{raj2023, Gong22}. Each of these torques plays a unique role in shaping the skyrmion trajectory. However, the dominant torque finally determines the trajectory and stability of skyrmion motion~\cite{kong2013, qin2022, yu21, raj2023, wang2020}. In insulating systems, magnons diffusing from hotter to cold regions carry negative angular momentum~\cite{kong2013, lin2014}, exerting an \textbf{adiabatic} STT that pushes skyrmions towards the hotter end, opposite the direction of magnon flow. In the insulator, Seebeck-effect-induced torque is negligible, and the dynamics are similar to the current-induced STT-driven motion, with both longitudinal and transverse velocity components governed by magnon STT~\cite{kong2013, lin2014}. Experimental studies on insulating helimagnets, such as $Cu_2OSeO_3$, confirm skyrmions' move towards the hotter end, validating the dominance of magnonic torques over other torque components. Both skyrmion velocity and Hall angle increase non-linearly with increases in $\boldsymbol{\nabla} T{\rm{(x)}}$. Note that the Hall angle is also independent of the skyrmion velocity, which differs from the skyrmions driven by the electric current that has been reported experimentally~\cite{Yu21}.

In metallic systems, the skyrmion response to a thermal gradient is more complex due to the coexistence of both magnonic and Seebeck-effect induced STT~\cite{raj2023, qin2022, Gong22}. Here, Seebeck-STT depends on the temperature differential across the racetrack and can oppose magnon-driven motion. Additionally, a temperature gradient induces spatial variation in saturation magnetization, exerting a dipolar field torque that pushes the skyrmion into the hotter region~\cite{raj2023, Gong22}. At higher temperatures, the magnon population is high, resulting in high entropy~\cite{raj2023, Yu21, Gong22}. This spatial entropy gradient along the thermal gradient gives rise to a non-adiabatic entropic torque that drives the skyrmion towards the cold end, where the entropy is lower~\cite{qin2022}. This non-adiabatic entropic force and Seebeck-STT act in opposition to magnonic and dipolar torques and can significantly influence the skyrmion trajectory, particularly under large temperature gradients. The net skyrmion trajectory thus results from the competition between magnonic, thermal, entropic, and dipolar torques, with the dominant mechanism dictated by material composition, device geometry, and the magnitude of $\boldsymbol{\nabla} T{\rm{(x)}}$~\cite{raj2023, Gong22}.  At low thermal gradients, magnonic torques often prevail, yielding hotter-end-directed motion; at higher gradients, thermal and entropic torques become significant and can reverse the direction of motion, propelling skyrmions towards the cold end while still exhibiting transverse (Hall-like) motion~\cite{qin2022}. Such behavior has been experimentally verified in metallic chiral magnets, notably in FeGe racetracks~\cite{qin2022}.

Previous studies have reported conflicting predictions regarding skyrmion motion in the presence of a temperature gradient. For instance, Gang \textit{et al.}~\cite{qin2022} and Gong {et al.}~\cite{Gong22} theoretically suggested that skyrmions can move either from the cold to the hotter end or in the opposite direction. This behavior is governed by the skyrmion’s spin configuration and material parameters, which collectively act to minimize the system’s magnetic energy~\cite{Raim2022, kong2013}.

The dynamics of skyrmions under a $\boldsymbol{\nabla} T{\rm{(x)}}$, have shown inconsistencies across theoretical and experimental studies~\cite{lin2014, qin2022, Yu21, raj2023, wang2020}. The fundamental mechanisms governing skyrmion trajectories remain unclear, particularly when considering variations in magnetic material parameters and the number of ferromagnetic (FM) layer, antiferromagnetic (AFM), and synthetic antiferromagnetic (SAF) layers~\cite{lin2014, qin2022, raj2023, wang18, Raim2022}. The question of whether skyrmions predominantly move towards the hotter or cold end in the presence of $\boldsymbol{\nabla} T{\rm{(x)}}$ remains an open question. Furthermore, the combined influence of electrical spin-transfer torque (STT) and thermal gradients on skyrmion motion has yet to be fully explored. Another crucial yet unresolved aspect is the thermal skyrmion Hall effect, which necessitates further investigation. Addressing these gaps requires a comprehensive understanding of skyrmion dynamics in thermally driven systems.

This study investigates skyrmion dynamics in a synthetic antiferromagnetic system across two sections. In the first section, we explore $\boldsymbol{\nabla} T(x)$-driven skyrmion dynamics, examining the effects of material parameters-damping constant ($\alpha$), Dzyaloshinskii-Moriya interaction ($D_{\text{int}}$), uniaxial anisotropy ($K_{\text{u}}$), and saturation magnetization ($M_s$)-under a fixed thermal gradient $\boldsymbol{\nabla} T(x)$. Our findings reveal thermally induced unidirectional motion of skyrmions towards the hotter region, accompanied by a negligible or small skyrmion Hall angle ($\theta_{\text{sk}}$). Both $\theta_{\text{sk}}$ and longitudinal skyrmion velocity ($V_{\text{x}}$) exhibit strong dependence on these parameters, with skyrmion size emerging as a critical factor in mediating magnon-induced dynamics.

\section{\label{sec:level2}Micro-magnetic Simulation}
We used the GPU-based micromagnetic simulation program Mumax3 to study skyrmion dynamics in a Co/Pt bilayer racetrack. The racetrack was modeled with dimensions $L_{\rm{x}}$$\times$$L_{\rm{xy}}$$\times$$L_{\rm{z}}$ = 256$\times$64$\times$1 $\rm{nm^3}$, which is divided into 200 slices of the dimension 1.28$\times$64$\times$1 $\rm{nm^3}$. A temperature gradient was established along the $+$x-direction by linearly increasing the temperature slice-by-slice while maintaining uniformity in the y and z directions, i.e., $\boldsymbol{\nabla} T{\rm(x)} \neq 0$ and $\boldsymbol{\nabla} T{\rm(y)}= \boldsymbol{\nabla} T{\rm(z)}$ = 0. The temperature is chosen below the Curie temperature ($T_{\rm{C}}$) to keep the system in the ferromagnetic phase at all times. The magnetic parameters are considered according to Sampio \textit{et} al.~\cite{sampaio13} for the Co/Pt racetrack. The material parameters, saturation magnetization $M_{\rm{s}} = 580 \times 10^{3}\,\rm{A/m}$, exchange stiffness constant $A_{\rm{ex}} = 15 \times 10^{-12}\,\rm{J/m}$, PMA constant $K_{\rm{u}}$=$0.8 \times 10^6\,\rm{J/m^3}$, dampin contant $\alpha$= 0.05 and interfacial-DMI (\(D_{\rm{int}}\)) = -$3\,\rm{mJ/m^2}$ for Co/Pt bilayer were taken from the experimental report~\cite{sampaio13}. The minimum energy state is achieved by numerically solving the Landau-Lifshitz-Gilbert (LLG) equation using the fourth-order Runge-Kutta method in finite difference discretization-based (open-source) software M{\scriptsize U}M{\scriptsize AX}3~\cite{Vansteenkiste14}. Here, the evolution of the space and time-dependent magnetization vector, \(\vec{m}(\vec{r}, t)\), is calculated at each cell of the discretized geometry by employing the following LLG equation-
\begin{equation}\label{eq: LLG}
\begin{aligned}
\frac{\partial\vec{m}}{\partial t} = 
\frac{\gamma_{\rm{LL}}}{1+\alpha^2}[\vec{m}\times(\vec{H}_{\rm{eff}})\\
+\alpha(\vec{m}\times(\vec{m}\times(\vec{H}_{\rm{eff}})))]+\vec{T}_{\rm{STT}}
 \end{aligned}
\end{equation}
where $\frac{\partial\vec{m}}{\partial t}$ is the Landau-Lifshitz torque, $\vec{m}$ is the reduced magnetization vector with respect to the saturation magnetization $M_{\rm{s}}$, $\gamma_{\rm{LL}}$ is the gyro-magnetic ratio (in rad/Ts), and $\vec{H}_{\rm{eff}}$ is the effective magnetic field.

At finite temperatures, magnetization fluctuates randomly due to energy exchange with the surrounding environment. This energy transfer leads to effects like lattice vibrations (phonons), moving electrons, and spin waves (magnons). To include these thermal effects in simulations, the stochastic field 
$\vec{H_{th}}$ is added to Eq. (1), with its autocorrelation given by~\cite{Vansteenkiste14}:
\begin{equation}\label{2}
\begin{aligned}
\langle \vec{H}_{th}^{ip}(t) \rangle=0\\
\langle \vec{H}_{th}^{ip}(t), \vec{H}_{th}^{jp}(t+\Delta t) \rangle=H_{th}\delta_{ij}\delta_{pq}\delta(\Delta t)
\end{aligned}
\end{equation}
where $\vec{H}_{th} = \vec{\eta}\sqrt{{\frac{2\alpha k_B T}{M_s \mu_0 \gamma_{LL} \Delta V \Delta t}}}$. The variable $\eta$ is treated as a random quantity following a Gaussian distribution with a mean of zero and a variance of $1$. $i$ and $j$ denote the micro-magnetic cells, and $p$, $q$
represent the cartesian components of the thermal field. $k_B$, $T$,
$V$, $t$, and $\delta$ are the Boltzmann constant, temperature, volume of each cell, time interval, and Dirac-delta function, respectively.
 
To account for the impact of thermal-induced STT, we incorporate both adiabatic and nonadiabatic STT into the stochastic LLG equation.~\cite{Leliaert2018, Vansteenkiste14}-
  \begin{equation}
 \label{3}
T_{STT} = 
a \vec{m}\times{\left[\frac{d\vec{m}}{dx}\times \vec{m} \right]-a\beta \left[\vec{m} \times \frac{\vec{dm}}{dx} \right]}
\end{equation}
where $a=\frac{\hbar \gamma_0 \rm{P} \rm{j(x)}}{2 \mu_0 e \rm{M}\rm{t_0}}$ is the STT coefficient induced by thermal current, $\beta$ is the non adiabatic STT factor and j(x) is the thermal gradient induced spin current which move towardss cold end. However, P, $\mu_0$, and $t_0$ are the spin polarization factor, absolute permeability, and the thickness of the magnetic layer, respectively.
\begin{figure}
\includegraphics[width = \linewidth]{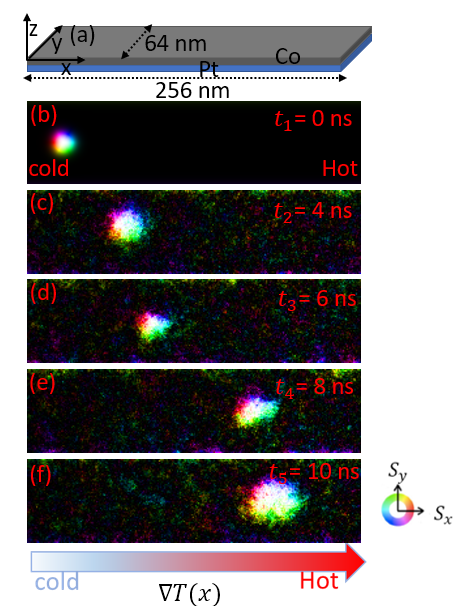}
\caption{\label{fig:1} (a) Schematic of the Co/Pt track. (b)-(f) Snapshot of the skyrmion position at different times.}
\end{figure}
Skyrmion dynamics under the $\boldsymbol{\nabla} T{\rm{(x)}}$ is primarily induced by several torque components that are defined as follows: 
\begin{equation}
\label{4}
 F_{\boldsymbol{\nabla} T(x)}= F_{\rm{\mu STT}}+F_{\rm{\tau STT}}+F_{\rm{ds}}+F_{\rm{DF}}
\end{equation}
where $F_{\rm{\mu STT}}$, $F_{\rm{\tau STT}}$, $F_{\rm{ds}}$, and $F_{\rm{DF}}$ are the magnonic torque, thermal torque, entropy torque, and thermally induced dipolar field torque, respectively~\cite {raj2023, Gong22}. Each component of these torques influences the dynamics of the skyrmion differently has been discussed. Therefore, the resulting skyrmion dynamics are due to the combined effects of these torque components. For example, $F_{\rm{\mu STT}}$ is induced by the propagation of magnon spin waves towards the colder regions, applying torque towards the hotter end, causing skyrmion movement in the hotter region which has been discussed~\cite{kong2013, raj2023}. $F_{\rm{\tau STT}}$ induced by the diffusion of electrons towards the colder side of the racetrack, where $F_{\rm{\tau STT}}$ pushes the skyrmion towardss the colder side of the racetrack. $F_{\rm{ds}}$ represents torque due to the entropy difference and depends on significant temperature differences, allowing the entropic force to act on the skyrmion, which is directed in the  $\boldsymbol{\nabla} T{\rm{(x)}}$ direction; hence, skyrmion motion is opposite to the applied thermal gradient direction. The last term, $F_{\rm{DF}}$, is due to the change in stray field into $\boldsymbol{\nabla} T{\rm{(x)}}$ direction. The local change in stray is due to the change in $M_s$ with the thermal gradient. The stray field gradient or torque due to the dipolar field pushes the skyrmion towardss the hotter side.\\

\begin{figure*}[hbt!]
\includegraphics[width = \textwidth]{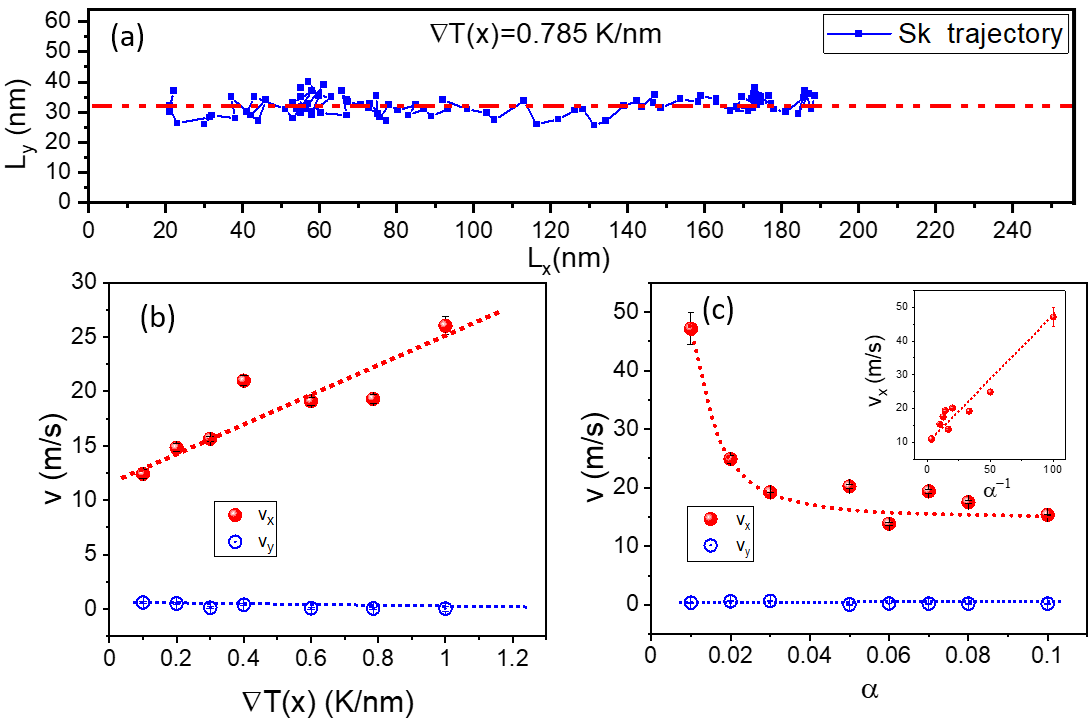}
\caption{\label{fig:2}(a) Representative skyrmion trajectory in racetrack for the $\boldsymbol{\nabla} T{\rm{(x)}}$=0.785 K/nm. (b) $v_{\rm{x}}$ and $v_{\rm{y}}$ as a function of $\boldsymbol{\nabla} T{\rm{(x)}}$. (c) $v_{\rm{x}}$ as a function of damping constant. Inset shows the $v_{\rm{x}}$ vs. $1/\alpha$. Dotted lines are guides to the eyes.}
\end{figure*}

\section{\label{sec:level2}Results and Discussion}
\subsection{Thermal gradient assisted skyrmion dynamics in FM layer}
We investigate the skyrmion dynamics in a ferromagnetic racetrack under the presence of $\boldsymbol{\nabla} T{\rm{(x)}}$. The results are organized in two parts: (i) the effect of varying $\boldsymbol{\nabla} T{\rm{(x)}}$ on skyrmion dynamics with fixed magnetic material parameters ($\alpha$, $D_{\rm{int}}$, $K_{\rm{u}}$, and $M_{\rm{s}}$); and (ii) effect of individual material parameters on skyrmion dynamics where $\boldsymbol{\nabla} T{\rm{(x)}}$ is fixed. We consider a Néel-type skyrmion in a Co/Pt racetrack (schematically shown at the top of Fig.~\ref{fig:1}(a)) at position x= 20 nm and y= 32 nm. A linear $\boldsymbol{\nabla} T{\rm{(x)}}$ = 0.785 K/nm is applied along the +x-direction, with the temperature increasing from 200 K to 400 K between $0 \leq \rm{x} \leq 256$ nm. 

As shown in the sequence of snapshots at different times in Fig.~\ref{fig:1} (b-e), the skyrmion position changes over time as it moves steadily towards the hotter end of the racetrack. Within a simulation time of 7 nanoseconds(ns), the skyrmion covers nearly 160 nm of the racetrack along the +x-direction, with negligible transverse displacement. This unidirectional motion indicates that dynamics are primarily governed by $\boldsymbol{\nabla} T{\rm{(x)}}$ induced torque. The unidirectional behavior is checked for different $\boldsymbol{\nabla} T{\rm{(x)}}$ values ranging from 0.1 to 1.2 K/nm.

A representative result for this behavior is shown for $\boldsymbol{\nabla} T{\rm{(x)}}$=0.785 K/nm in Figure~\ref{fig:2}(a), where we plot the trajectory of the skyrmion along the length of the track. The skyrmion center was extracted from OVF files using a MATLAB-based code. At $t = 0$ ns, it is located at $x = 20$ nm and $y = 32$ nm, where the local temperature is 212 K in this case. The trajectory confirms steady motion towards the hotter end of the racetrack, with negligible transverse deflection. Interestingly, the skyrmion position fluctuates slightly surrounding the racetrack center, as seen in Fig.~\ref{fig:2}(a). This behavior is attributed to fluctuation in skyrmion size caused by the temperature gradient, which affects magnetic material parameters such as $K_{\rm{u}}$, $D_{\rm{int}}$, and $M_{\rm{s}}$. As a result, the skyrmion does not behave like a rigid particle as earlier reported \cite{Reichhardt17, Mishra25} but exhibits size-dependent wobbling. Minor transverse displacements arise from Brownian motion and stochastic thermal fluctuations. While most prior studies report a thermally driven skyrmion Hall effect under $\boldsymbol{\nabla} T(x)$ \cite{raj2023, kong2013, qin2022, Raim2022, Yu21}, our analysis reveal that the Hall angle is suppressed and remains nearly zero. However, Z. Wang \textit{et al.} demonstrated skyrmion motion without transverse diffusion from hotter to cold regions, attributed to the interplay of repulsive forces between skyrmions, thermal spin-orbit torques, entropic forces, and magnonic spin torques~\cite{wang2020}. Notably, the detailed investigation of the thermally induced skyrmion Hall effect in purely $\boldsymbol{\nabla} T(x)$-driven skyrmion motion from cold to hot regions has not been reported previously. Thus, our work highlights the efficiency of thermally induced torque in driving skyrmions along nearly straight paths. In the following sections, we provide a more detailed analysis of the torque mechanisms responsible for this behavior.

In our analysis, skyrmions move towards the hotter end without exhibiting a thermal-induced skyrmion Hall effect (t-SHE), driven primarily by a net torque induced by the temperature gradient, $\boldsymbol{\nabla} T(x)$, as described by eq.~\ref{4}. This motion suggests that the combined torques, $F_{\mu \text{STT}} + F_{\text{DF}}$, dominate over $F_{\tau \text{STT}} + F_{\text{ds}}$. To understand the underlying mechanisms in our system, it is essential to decouple and independently control the contributions of $F_{\mu \text{STT}}$ and $F_{\text{DF}}$ through systematic studies of individual magnetic parameters. Previous studies and magnon spectrum analyses~\cite{Szulc21, kong2013} indicate that $F_{\mu \text{STT}}$ can be tuned by adjusting the Gilbert damping parameter $\alpha$, the temperature gradient $\boldsymbol{\nabla} T(x)$, the uniaxial anisotropy constant $K_{\text{u}}$, and the Dzyaloshinskii-Moriya interaction strength $D_{\text{int}}$, whereas $F_{\text{DF}}$ primarily depends on the saturation magnetization $M_s$. To investigate the origins of this remarkable skyrmion dynamics, we systematically explore the effects of these magnetic parameters. Furthermore, we examine the role of skyrmion size, modulated by variations in intrinsic magnetic parameters, as a critical determinant of SHE-free skyrmion dynamics- a novel finding not previously reported.

To investigate the influence of the thermal gradient on skyrmion dynamics, we systematically varied the value of $\boldsymbol{\nabla} T(x)$ over a range of 0.1 to 1.2 K/nm. The increases of $\boldsymbol{\nabla} T(x)$ enhances the magnon spin current, $j_{\text{x}}$, which is defined as:
\begin{equation}
\vec{j_x}=\frac{\pi}{24}a^2(\frac{k_B}{\hbar s})^2\frac{T}{\alpha}\boldsymbol{\nabla}T(x)
\label{6}
\end{equation}
where a, $k_B$, $s$, $T$ are the lattice constant, Boltzmann constant, spin, and temperature, respectively~\cite{kong2013}. The simulations were performed using fixed material parameters for the Co/Pt racetrack, as detailed in the section~\uppercase{ii}, the geometric parameters used as Fig.~\ref{fig:1}. Figure~\ref{fig:2}(b) presents the mean skyrmion velocity components (longitudinal ($v_{\text{x}}$) and transverse ($v_{\text{y}}$)), as a function of $\boldsymbol{\nabla} T(x)$. Each data point represents the average of eight independent simulations, with consistent procedures applied throughout. The skyrmion position as a function of time was extracted from OVF files using MATLAB code.

Our results demonstrate that the $v_{\text{x}}$, increases linearly with the $\boldsymbol{\nabla} T(x)$, while the transverse velocity, $v_{\text{y}}$, remains negligible, as shown in Fig.~\ref{fig:2}(b). This linear increase in $v_{\text{x}}$ supports the notion that a higher temperature gradient enhances the magnon spin current, as described by eq.~\ref{6}, thereby increasing the driving torque and skyrmion velocity. A higher Gilbert damping constant, $\alpha$, reduces magnon lifetimes, decreasing the magnon spin current, $\vec{j_{\text{x}}}$, as described by eq.~\ref{6}, thus enabling indirect control over magnon dynamics~\cite{kong2013}. Fujita \textit{et al.} experimentally determined that the lowest value of $\alpha$ for Co/Pt multilayers, measured using ferromagnetic resonance (FMR), is approximately 0.02 at an optimal Co layer thickness of 1.8 nm~\cite{Fujita2019}. We systematically varied $\alpha$ over a range of 0.01 to 0.1. Figure~\ref{fig:2}(c) shows that $v_{\text{x}} \propto 1/\alpha$ (see inset of Figure~\ref{fig:2}(c)), while variation of $v_{\text{y}}$ remains negligible across all values of $\alpha$. In both cases ($\boldsymbol{\nabla} T(x)$- and $\alpha$-dependent skyrmion dynamics), $v_{\text{x}}$ aligns with theoretical expectations~\cite{kong2013}. In contrast, the negligible variation of $v_{\text{y}}$ contradicts from prior theoretical predictions.

The $\boldsymbol{\nabla} T(x)$ induces the longitudinal skyrmion velocity, $v_{\text{x}}$, described by the following relation:
\begin{equation}
    \vec{v_{\rm{x}}}=\gamma_{LL} J a^2\vec{j_x}-\frac{\gamma_{LL} \eta\alpha a^2 k_B}{\pi Q^2}\boldsymbol{\nabla} T(x)=\vec{v}_{\rm{x}}^M-\vec{v}^B
    \label{7}
\end{equation}
where $\vec{v}_{\text{x}}^M$ and $\vec{v}^B$ represent the magnonic and Brownian velocity components, respectively. The $\vec{v}_{\text{x}}^M$ drives the skyrmion towards the hotter side, while the $\vec{v}^B$ drives it towards the cold side. The parameters $a$, $\gamma_{\text{LL}}$, $\eta$, $Q$, $\vec{j_x}$, and $J$ denote the lattice constant, gyromagnetic ratio, dimensionless coefficient, topological charge, magnon spin current, and exchange coupling constant, respectively~\cite{kong2013}. In our system, skyrmions move towards the hotter side, indicating the dominance of the $\vec{v}_{\text{x}}^M$ over $\vec{v}^B$ in our case. The material-dependent magnonic velocity, $\vec{v}_{\text{x}}^M$, is given by:
\begin{equation}
    \vec{v}_{\rm{x}}^M=\frac{\gamma_{LL} \eta a^2 J k_B T}{6D_{int}^2 Q^2 \alpha}\boldsymbol{\nabla} T(x)
    \label{8}
\end{equation}
Additionally as per reference~\cite{kong2013}.The  $\vec{v}_{\rm{y}}$ is related to $v_{\text{x}}^M$ as follows:
\begin{equation}
    \vec{v}_{\rm{y}}=2\alpha\eta \vec{v}_{\rm{x}}^M
    \label{vy}
\end{equation}
indicating that $v_{\text{y}}$ also depends on $\alpha$ and $v_{\text{x}}^M$~\cite{kong2013}.

Interestingly, as shown in Fig~\ref{fig:2}(b) \& (c), we find $v_{\text{x}}$$\propto\boldsymbol{\nabla} T{\rm{(x)}}$ and $v_{\text{x}}\propto 1/\alpha$, consistent with eqs.~\ref{7} \& ~\ref{8}. However, Fig.~\ref{2}(a-c) clearly show that the $v_{\text{y}}$ component of the skyrmion nearly zero and insensitive to both $\alpha$ and $\boldsymbol{\nabla} T{\rm{(x)}}$, in contrast to the prediction by Kong \textit{et al.}~\cite{kong2013}. This discrepancy suggests that the dependence of $v_{\text{y}}$ on the temperature gradient may not be universal. Our results for a Co/Pt bilayer show that existing models do not fully capture the combined effects of the spin-transfer torque force $F_{\mu \text{STT}} $ and the damping force $F_{\text{DF}} $. To investigate this further, we systematically tuned $F_{\mu \text{STT}}$ by varying magnetic parameters such as $D_{\text{int}}$ and the uniaxial anisotropy constant, $K_{\text{u}}$, and controlled $F_{\text{DF}}$ by adjusting the saturation magnetization, $M_{\text{s}}$, which reduces magnon flow but enhances the dipolar field-induced torque. Throughout this study, we maintained a fixed $\boldsymbol{\nabla} T{\rm{(x)}}$ and varied only one magnetic parameter at a time.

\begin{figure*}[hbt!]
\includegraphics[width = \textwidth]{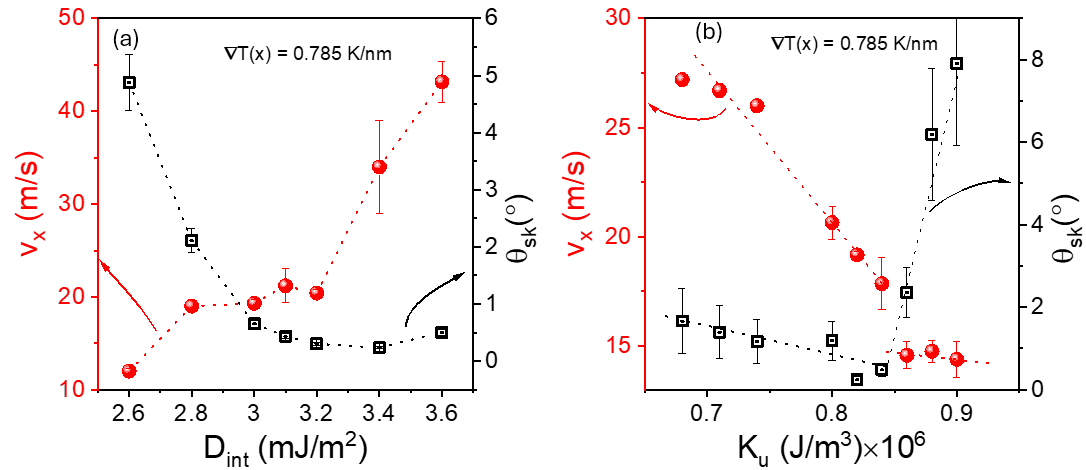}
\caption{\label{fig:3}(a) $v_{\rm{x}}$ and $\theta_{\rm{sk}}$ as a function of $D_{\rm{int}}$ at a fixed thermal gradient $\boldsymbol{\nabla} T_{\rm{x}} = 0.785$ K/nm. (b) $v_{\rm{x}}$ and $\theta_{\rm{sk}}$ as a function of $K_{\rm{u}}$ at the same gradient. In both panels, the red symbols represent the $v_{\rm{x}}$ component, while the black symbols represent $\theta_{\rm{sk}}$. Dotted lines are guides to the eyes.}
\end{figure*}
The magnon-induced skyrmion velocity is influenced by the $D_{\text{int}}$ as predicted theoretically by eq.~\ref{8}. However, this effect has not been previously reported for metallic systems. To investigate the role of $D_{\text{int}}$ in our system, we varied $D_{\text{int}}$ over a range of 2 to 4 mJ/m$^2$, within which skyrmion is found to be energetically stable during motion. Beyond this range, the skyrmion annihilated (supplementary Figure~SF2). A. Cao \textit{et al.} experimentally reported a high value of $D_{\text{int}} = 3.3$ mJ/m$^2$ in Pt/Co/MgO multilayer samples~\cite{Cao_2020}. Figure~\ref{fig:3}(a) presents the $v_{\text{x}}$ and skyrmion Hall angle ($\theta_{\text{sk}}=tan^{-1}(\frac{v_y}{v_x})$ as functions of $D_{\text{int}}$, evaluated at a fixed $\boldsymbol{\nabla} T(x) = 0.785$ K/nm, with all other magnetic parameters held constant, as described in Section~\uppercase{ii}. We observed the skyrmion dynamics within the  $D_{\text{int}}$ range of 2.6 to 3.6 mJ/m$^2$, two distinct dynamical regimes emerge Fig~\ref{fig:3}(a).

For an $2.6 \leq D_{\text{int}} \leq 2.8 , \text{mJ/m}^2$, $v_x$ increases by approximately 100\%, from 10 to 20 m/s, while the $\theta_{\text{sk}}$ decreases from $\sim$5° to $\sim$2° (Fig.~\ref{fig:3}(a)). Specifically, at $D_{\text{int}} = 2.6~\text{mJ/m}^2$, analysis of the magnetization profile reveals a small skyrmion size, with complete annihilation occurring around 5 ns during the $\boldsymbol{\nabla} T(x)$ induced motion, as shown in the trajectory of skyrmion in left pannel of supplementary Figure~SF2 (Supplementary Section SR-II). In the range $2.8 \leq D_{\text{int}} \leq 3.2 , \text{mJ/m}^2$, $v_x$ remains nearly constant, while $\theta_{\text{sk}}$ further decreases from approximately 2° to 0.5° (Fig.~\ref{fig:3}(a)). For $D_{\text{int}} \geq 3.2 , \text{mJ/m}^2$, $v_x$ increases linearly up to 45 m/s, while $\theta_{\text{sk}}$ stabilizes at approximately 0.5° as a function of $D_{\text{int}}$ (Fig.~\ref{fig:3}(a)). At $D_{\text{int}} = 3.6$ mJ/m$^2$, the skyrmion shape becomes elongated along the $x$-axis (towards the hotter side) during motion. After 1.7 ns, the skyrmion elongation makes it difficult to determine its exact center. This behavior is shown in the right panel of Supplementary Figure~SF2 (see Supplementary Section SR-II). Consequently, fluctuations in the $y$-direction increase, as evidenced by the trajectory of skyrmion in Supplementary Figure SF2 (right side pannel) and thefore $\theta_{\text{sk}}$ is sightly large (Fig.~\ref{fig:3}(a)).These trend in $v_x$ exhibits two distinct transitions at critical DMI values of  $D_{\text{int}} = 2.8 , \text{mJ/m}^2$ and $D_{\text{int}} = 3.2 , \text{mJ/m}^2$, as observed in Fig.~\ref{fig:3}(a). A similar trend is observed for the difefrent value of $\boldsymbol{\nabla} T(x) = 0.5, \text{K/nm}$ (see Supplementary Section SR-III), confirming the robustness of these findings. Notably, these $D_{\text{int}}$-dependent results deviate from the predictions of eq.~\ref{8}. The $D_{\text{int}}$-dependent skyrmion dynamics discussed here (Fig.~\ref{fig:3}(a)) can be interpreted through the magnon dispersion relation, as discussed below:

\begin{equation}
    \omega=\frac{\gamma_{LL}}{\mu_{0}M_{sat}}[2A_{ex}k^2 \pm 2D_{int}k+2K_u]
    \label{10}
\end{equation}
\begin{equation}
   v_g=\frac{d\omega}{dk} =\frac{\gamma_{LL}}{\mu_{0}M_{sat}}[4A_{ex}k \pm 2D_{ind}]
    \label{11}
\end{equation}
Equation~\ref{10} describes the magnon dispersition as a function of $D_{\rm{int}}$, $M_{\rm{sat}}$ and $K_{\rm{u}}$. As predicted, increasing $D_{\rm{int}}$ enhances the nonreciprocal propagation of magnons due to chiral interactions~\cite{Moon13,  Wang16, Szulc21}. This nonreciprocity leads to directional asymmetry in the magnon group velocity: $D_{\rm{int}}$ increases the magnon velocity along the direction of the $D_{\rm{int}}$-induced effective field while reducing it in the opposite direction. Eq.~\ref{11}, provides the group velocity $v_g$ as a function of the exchange stiffness ($A_{\rm{ex}}$), $M_{\rm{sat}}$ and $D_{\rm{int}}$. The $\boldsymbol{\nabla} T(x)$ induced magnon spin current density ($\vec{J}_m$) can be expressed mathematically by integrating over all possible magnon momenta in k-space. The expression for $\vec{J}_m$ in the case of a ferromagnetic insulator (FMI) and non-magnetic heterostructures is given as follows:
\begin{equation}
   \vec{J^x_m}=-\frac{\hbar}{(2\pi)^3}\int d^3k~\tau_k~\frac{\partial{n^0_k}}{\partial{T}}\vec{v_g}(\vec{v_g}.\boldsymbol{\nabla} T(x))
    \label{12}
\end{equation}
where, $\tau_k$ is the $k$-magnon relaxation time, ${v}_g$ is the magnon group velocity, $n^0_k$ is the equilibrium magnon distribution, given by the Bose-Einstein distribution\cite{Rezende14, Rezende16}. The negative sign in right hand side indicate the $ \vec{J^x_m}$ opposite to $\boldsymbol{\nabla} T(x)$.

Figure~\ref{fig:3}(a) shows that as $v_x$ increases, $\theta_{\text{sk}}$ and $v_y$ decrease (for the $v_y$ trend, see the right panel of Supplementary Figure~SF3 in Section SR-III). This behavior differs from electrically driven spin-transfer torque (STT) skyrmion dynamics, where $\theta_{\text{sk}}$ increases with $v_x$ or with the electrical current density $j_e$~\cite{Jiang16}. Across the range of $D_{\text{int}}$, $v_x$ increases in discrete steps, while $\theta_{\text{sk}}$ is continuously decreases and follows a trend similar to $v_y$. These findings reveal significant deviations from theoretical predictions when a thermal gradient $\boldsymbol{\nabla}T(x)$ drives skyrmion motion. In particular, the model proposed by Kong \textit{et al.}~\cite{kong2013} does not capture the $D_{\text{int}}$-dependent skyrmion dynamics under $\boldsymbol{\nabla}T(x)$. It fails to predict the stepwise enhancement of $v_x$ with $D_{\text{int}}$, which governs the onset of pronounced skyrmion motion. Similarly, the magnon current density $\vec{J_m}$ (eq.~\ref{12}) does not account for this stepwise variation in $v_x$. Overall, our results indicate that $v_x$ and $\theta_{\text{sk}}$ remain independent of each other under a thermal gradient, since $\theta_{\text{sk}}$ does not exhibit step-like behavior.

An increase in the uniaxial anisotropy constant ($K_{\text{u}}$) widens the magnon bandgap, thereby reducing the magnon group velocity, ${v}_{\text{g}}$, as described by eq.~\ref{11} and~\ref{10}. This reduction arises due to the stronger alignment of spins along the $z$-direction. Although ${v}_{\text{g}}$ is not directly a function of $K_{\text{u}}$, but $K_u$ influences other $k$-dependent terms, affecting the mode character, boundary conditions, dipolar interactions, and the magnon bandgap~\cite{Delgado24, Yimer23}. To study the impact of $K_{\text{u}}$ on skyrmions, we evaluated $v_x$ and $\theta_{\text{sk}}$ as a function of $K_{\text{u}}$ at a fixed thermal gradient $\boldsymbol{\nabla} T(x) = 0.785$ K/nm, with all other parameters as described in Section~\uppercase{ii} held constant. Skyrmions remain stable during motion in the range $0.68 \times 10^6 \leq K_{\text{u}} \leq 0.9 \times 10^6$ J/m$^3$. For example, at $K_{\text{u}} = 0.68 \times 10^6$ J/m$^3$, the skyrmion deforms and annihilates after 5.5 ns during motion see the left pannel in supplementary Figure~SF4 (Supplementary Section SR-IV). At $K_{\text{u}} = 0.92 \times 10^6$ J/m$^3$, the skyrmion also annihilates during motionsee the right pannel in supplementary Figure~SF4 (Supplementary section SR-IV). Therefore, we restrict our analysis to the range $0.68 \times 10^6 \leq K_{\text{u}} \leq 0.9 \times 10^6$ J/m$^3$. Experimentally, the highest reported value of $K_{\text{u}}$ is $0.8 \times 10^6$ J/m$^3$~\cite{sampaio13}. We observe that both $v_{\text{x}}$ and $\theta_{\text{sk}}$ exhibit systematic changes with $K_{\text{u}}$. For $0.68 \times 10^6 \leq K_{\text{u}} \leq 0.84 \times 10^6$ J/m$^3$, $v_{\text{x}}$ decreases linearly by about 50\%, from $\sim$28 m/s to $\sim$15 m/s, while $\theta_{\text{sk}}$ also decreases linearly from $\sim$1.66° to $\sim$0.46°. For $K_{\text{u}} \geq 0.84 \times 10^6$ J/m$^3$, $v_{\text{x}}$ saturates at $\sim$15 m/s, whereas $\theta_{\text{sk}}$ increases linearly with $K_{\text{u}}$ at much higher rate of $\frac{d\theta_{\rm{sk}}}{dK_{\rm{u}}}$ (Fig.~\ref{fig:3}(b)). The change in $\theta_{\text{sk}}$ is consistent with that of $v_y$ see in the supplementary Figure~SF5 (see Supplementary Section SR-V).

This behavior shows a clear threshold at $K_{\text{u}} = 0.84 \times 10^6$ J/m$^3$, separating two distinct dynamical regimes. For $K_{\text{u}} < 0.84 \times 10^6$ J/m$^3$, $v_{\text{x}}$ decreases linearly with increasing $K_{\text{u}}$, consistent with the magnon dispersion relation as described in eq.~\ref{10}, which shows that the magnon band-gap increases with $K_{\text{u}}$, resulting in reduced $v_g$. Therefore, enhancement in $K_{\text{u}}$ may indirectly reduces the  $\vec{J^x_m}$ (eq.~\ref{12}). In this range, $\theta_{\text{sk}}$ also decreases linearly with $K_{\text{u}}$, although its magnitude remains much smaller compared to STT-driven or other reported $\boldsymbol{\nabla} T(x)$-driven skyrmion dynamics~\cite{Yimer23, Litzius17}. Above the threshold, $v_{\text{x}}$ saturates at $\sim$15 m/s, while $\theta_{\text{sk}}$ increases linearly up to $\sim$8°, a trend not explained by the magnon dispersion relation or by $\vec{J^x_m}$ (eqs.~\ref{10} and~\ref{12}). These dynamics deviate from the theoretical predictions of eqs.~\ref{10} and~\ref{12}, revealing a regime not captured by conventional theory.

So far, we controlled the magnon spin current ($\vec{j_{\text{x}}}$ or $\vec{J^x_m}$) by varying $D_{\text{int}}$, $K_{\text{u}}$, $\alpha$, and $\boldsymbol{\nabla} T(x)$. The $v_{\text{x}}$ shows systematic variation only with $\alpha$ and $\boldsymbol{\nabla} T(x)$ which are consistent with theoretical calculations~\cite{kong2013}. In contrast, $v_{\text{y}}$ shows negligible dependence on both $\alpha$ and $\boldsymbol{\nabla} T(x)$, a behavior not reported in literature previously. For $D_{\text{int}}$ and $K_{\text{u}}$ dependent analysis, the skyrmion dynamics are markedly different. In both cases, $v_{\text{x}}$ and $\theta_{\text{sk}}$ are restricted by a critical value, as discussed above. For example, in the case of $D_{\text{int}}$, when $D_{\text{int}} > 3.2$ mJ/m$^2$, $v_{\text{x}}$ increases rapidly with $D_{\text{int}}$ while $\theta_{\text{sk}}$ remains nearly constant at near to zero. Similarly, for $K_{\text{u}} < 0.84 \times 10^6$ J/m$^3$, the $v_{\text{x}}$ decreases with $K{\text{u}}$, and $\theta_{\text{sk}}$ again stays close to zero. In contrast, in both cases, when $v_{\text{x}}$ becomes nearly constant, $\theta_{\text{sk}}$ changes rapidly. Therefore, both $\theta_{\text{sk}}$ and $v_{\text{x}}$ exhibit significant variations beyond the critical values of $D_{\text{int}}$ and $K_{\text{u}}$ (Fig.~\ref{fig:3}). Thus, $\theta_{\text{sk}}$ varies independently of $v_{\text{x}}$, and this behavior is not explained by existing theoretical models~\cite{kong2013}.

\begin{figure}
\includegraphics[width=\linewidth]{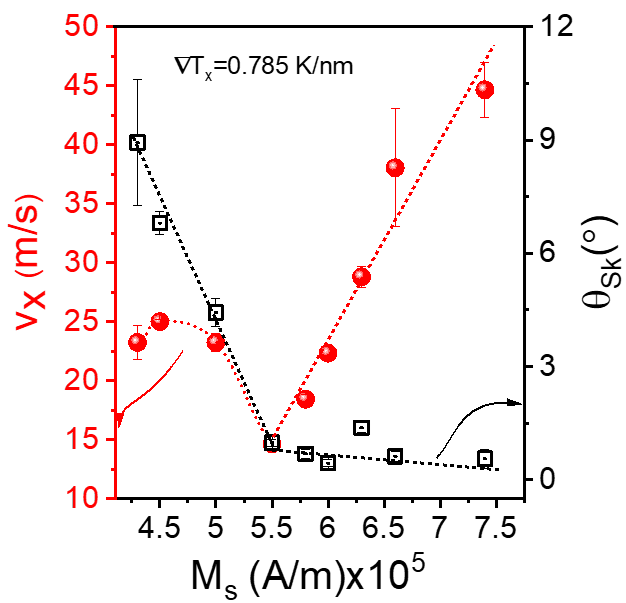}
\caption{\label{fig:4} $v_{\rm{x}}$ and $\theta_{\rm{sk}}$ as functions of $M_{\rm{s}}$ at a fixed thermal gradient $\boldsymbol{\nabla} T_{\rm{x}} = 0.785$ K/nm. Red symbols represent the $v_{\rm{x}}$} component, while black symbols represent $\theta_{\rm{sk}}$.
\end{figure}

Next, we explore skyrmion dynamics by varying the saturation magnetization, $M_{\text{s}}$, which influences the dipolar field induced torque. According to eq.~\ref{11}, the group velocity $v_{g} \propto D_{\text{int}}/M_{s}$ and $A_{\text{ex}}/M_{s}$. Thus, increasing $M_{s}$ reduces the DMI-induced nonreciprocal contribution (for fixed $D_{\text{int}}$). Similarly, an increase in $M_{s}$ also suppresses the exchange contribution to $v_{g}$ at fixed $A_{\text{ex}}$. To investigate this and check the dipolar field torque effect on skyrmion dynamics, we systematically varied $M_{\text{s}}$ within the skyrmion stability range during the motion $4.3 \times 10^5 \leq M_{\text{s}} \leq 7.4 \times 10^5$ A/m, while fixing $\boldsymbol{\nabla} T(x) = 0.785$ K/nm and keeping all other parameters ($D_{\text{int}}$, $K_{\text{u}}$, and $\alpha$) constant, as described in Section~\uppercase{ii}. We extracted $v_x$, $v_y$, and $\theta_{\text{sk}}$ as functions of $M_{\text{s}}$ (Fig.~\ref{fig:4}). For $M_{\text{s}} < 4.3 \times 10^5$ A/m, the skyrmion moves towards the cold side (Supplementary Fig.~SF6, left panel, Section SR-VI). In contrast, at $M_{\text{s}} = 7.4 \times 10^5$ A/m, the skyrmion moves towards the hotter side, continuously deforms, and eventually annihilates (Supplementary Fig.~SF6, right panel, Section SR-VI). Increasing $M_{\text{s}}$ reduces $\vec{J^x_m}$ ( since $\propto v_g \propto 1/M_{s}$), thereby weakening the magnon-induced torque, while simultaneously enhancing the dipolar-field-induced torque due to a stronger stray-field gradient under $\boldsymbol{\nabla} T(x)$. As a result, we observe motion towards the hotter region with random fluctuations in both the $x$- and $y$-directions, consistent with the trends found in the $D_{\text{int}}$ and $K_{\text{u}}$ studies. For $4.3 \times 10^5 \leq M_{\text{s}} \leq 5.5 \times 10^5$ A/m, $v_{\text{x}}$ decreases from $\sim$25 m/s to $\sim$15 m/s, while $\theta_{\text{sk}}$ decreases from $\sim$7° to $\sim$1°(Fig.~\ref{fig:4}). For $M_{\text{s}} \geq 5.5 \times 10^5$ A/m, $v_{\text{x}}$ increases linearly from $\sim$15 m/s to $\sim$45 m/s, while $\theta_{\text{sk}}$ $\sim$~0.5°(Fig.~\ref{fig:4}). Notably, $\theta_{\text{sk}}$ follows $v_y$ (Supplementary Fig.~SF7) and remains independent of $v_x$. This non-monotonic trend highlights the competition between thermal magnonic forces and dipolar field torque, offering insight into the tunability of skyrmion transport under controlled thermal gradients in spintronic devices. Thus, the $M_{\text{s}}$-dependent analysis indicates that above a critical value of $M_{\text{s}}$, skyrmion motion is primarily driven by the thermal gradient–induced dipolar field torque. In this regime, $\theta_{\text{sk}}$ becomes nearly constant and approaches zero.  

In above studies we investigated skyrmion dynamics driven independently by the magnon spin current-induced torque and the dipolar field-induced torque. As discussed above, the $\theta_{\text{sk}} $ remains negligible (0.5°–2°) in certain ranges  of material parameters for both cases. Earlier reported results on $\theta_{\text{sk}} $ induced skyrmion dynamics showed $\theta_{\text{sk}} $ is in the range of $8^{\circ}$-$18^{\circ}$ \cite{qin2022, Yu21}. Our results $v_{\rm{x}}$ vs. $\theta_{\text{sk}} $ for the all cases shows the no correlation  in contrary to the predictions given by eq.\ref{vy}~\cite{kong2013}. We find that when $v_{\text{x}}$ increases, variation of $\theta_{\text{sk}}$ remains negligible. This is in contrast to STT- or SOT-induced skyrmion dynamics, in which $\theta_{\text{sk}} \propto v_{\text{x}}$~\cite{Jiang13, Litzius17}. Finally, we investigate the influence of skyrmion size on temperature-gradient-induced skyrmion dynamics. Recently, Zeissler \textit{et al.} reported a diameter-independent skyrmion Hall angle in magnetic multilayers~\cite{Zeissler20}, identifying a size range of 35–85 nm where the skyrmion Hall angle reaches a minimum of approximately $11^{\circ}$. To develop a deeper understanding of how intrinsic material parameters affect skyrmion size and its velocity during motion along the racetrack, we conduct a detailed systematic analysis. For STT/SOT-driven dynamics, the theoretical $\theta_{\text{sk}}$ for a rigid skyrmion is:
\begin{equation}
\theta_{sk} \approx \frac{\pm 8\Delta}{\alpha \pi^2 R}
\label{13}
\end{equation}
where $\Delta = \sqrt{A / K_{\text{eff}}} = 4.33$ nm is the domain wall width, $R$ is the skyrmion diameter, $\alpha$ is the Gilbert damping constant, and $A$ is the exchange stiffness~\cite{Jiang13}. However, experimental reports indicate a much smaller $\theta_{\text{sk}}$, often independent of skyrmion size in current-driven skyrmion dynamics~\cite{Zeissler20}. Liu \textit{et al.} predicted a vanishing $\theta_{\text{sk}}$ for elongated skyrmions, suggesting $\theta_{\text{sk}}$ manipulation via skyrmion deformation~\cite{Liu25}. These findings indicate that $v_{\text{x}}$ and $\theta_{\text{sk}}$ are strongly correlated with skyrmion size and shape.

\begin{figure*}[hbt!]
\includegraphics[width = \textwidth]{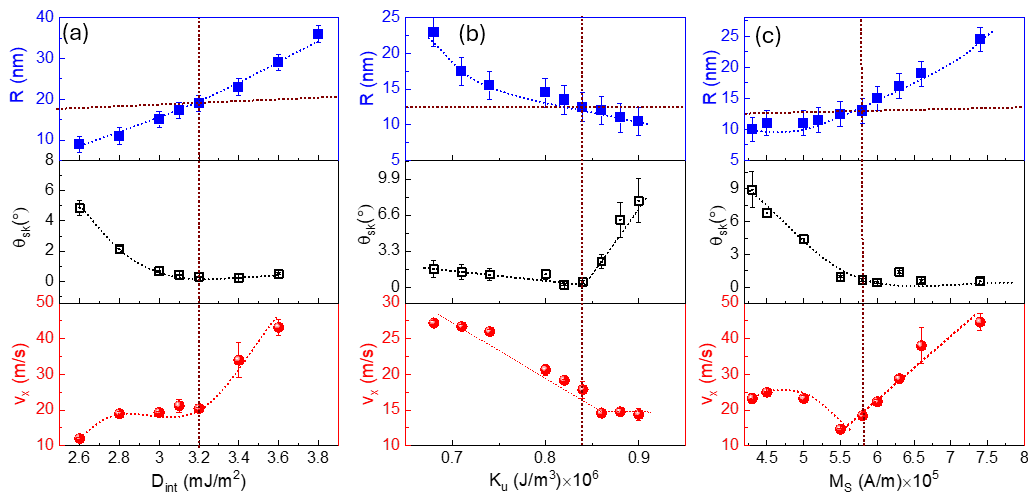}
\caption{\label{fig_dia:5} Variation of skyrmion diameter ($R$), $\theta_{\rm sk}$, and $v_{x}$ (top to bottom) as a function of (a) $D_{\rm int}$, (b) $K_{u}$, and (c) $M_{s}$. Dotted lines are guides to the eyes. Vertical lines indicate the relevant material-parameter thresholds, and horizontal dotted lines mark the corresponding skyrmion-diameter at thresholds.}
\end{figure*}

 The skyrmion diameter $R$ has been extracted as a function of these parameters at $T = 0$ by analyzing the magnetization profile along the $x$-axis and determining the domain-wall length using Python-based data processing. We then plotted $R$ as a function of material parameters together with the corresponding $v_{\text{x}}$ and $\theta_{\text{sk}}$ values, establishing their correlation with skyrmion size (Fig.~\ref{fig_dia:5}). Notably, when $\theta_{\text{sk}}$ is calculated in our case from eq.~\ref{13}, it is significantly larger than that obtained from $\boldsymbol{\nabla}T(x)$-driven dynamics, deviating by nearly $70^{\circ}$–$80^{\circ}$. This discrepancy suggests the presence of additional factors influencing the observed dynamics.

Top to bottom in Fig.~\ref{fig_dia:5}(a) show $R$, $v_{\text{x}}$, and $\theta_{\text{sk}}$ as functions of $D_{\text{int}}$, respectively. We observe that $R$ increases almost linearly with $D_{\text{int}}$. Correlating $R$ with $v_{\text{x}}$ and $\theta_{\text{sk}}$ reveals that for $D_{\text{int}} > 3.2$ mJ/m$^2$, $v_{\text{x}}$ increases linearly while $\theta_{\text{sk}}$ is constant and close to zero. At this value of $D_{\text{int}}$, the skyrmion diameter is $R \sim 19$ nm (indicated by the dotted lines cross point), suggesting that beyond this size, $\theta_{\text{sk}}$ becomes minimal or nearly zero. Similarly, Fig.~\ref{fig_dia:5}(b) shows $R$, $v_{\text{x}}$, and $\theta_{\text{sk}}$ as functions of $K_{\text{u}}$, where $R$ decreases with increasing $K_{\text{u}}$. For $K_{\text{u}} < 0.84 \times 10^6$ J/m$^3$, $v_{\text{x}}$ decreases with $K_{\text{u}}$, while $\theta_{\text{sk}}$ remains close to zero, identifying $K_{\text{u}} = 0.84 \times 10^6$ J/m$^3$ as the critical value. For $M_{\text{s}}$ dependent analysis, the extracted values of $R$, $v_{\text{x}}$, and $\theta_{\text{sk}}$ are shown in Fig.~\ref{fig_dia:5}(c) (top to bottom). Here, $R$ increases with $M_{\text{s}}$. Correlating $R$ with $v_{\text{x}}$ and $\theta_{\text{sk}}$ shows that for $M_{\text{s}} > 5.5 \times 10^5$ A/m, $v_{\text{x}}$ increases linearly while $\theta_{\text{sk}}$ becomes nearly constant at zero.

\begin{table}[htbp]
\centering
\caption{Dependence of skyrmion Hall angle ($\theta_{\mathrm{sk}}$) and diameter on material parameters.}
\begin{tabular}{|l|c|}
\hline
\textbf{Material Parameters} & $\boldsymbol{\theta_{\mathrm{sk}}}$ \\
\hline
\begin{tabular}[c]{@{}l@{}}%
$3~\mathrm{mJ/m^2}\leq$$D_{\mathrm{int}} \leq 3.6~\mathrm{mJ/m^2}$\\
$0.68\times 10^6~\mathrm{J/m^3}\leq$$K_{\mathrm{u}} \leq 0.84\times 10^6~\mathrm{J/m^3}$\\
$5.5 \times 10^5 ~\mathrm{A/m} \leq M_{\text{s}} \leq 7.4 \times 10^5 ~\mathrm{A/m}$
\end{tabular}
& $0.5^{\circ}\leq\theta_{\mathrm{sk}}$$\leq 2^{\circ}$  \\[6pt]
 \hline
\begin{tabular}[c]{@{}l@{}}%
$2.6~\mathrm{mJ/m^2}\leq$$D_{\mathrm{int}} \leq 3~\mathrm{mJ/m^2}$\\
$0.84\times 10^6~\mathrm{J/m^3}\leq$$K_{\mathrm{u}} \leq 0.9\times 10^6~\mathrm{J/m^3}$\\
$4.3 \times 10^5 ~\mathrm{A/m} \leq M_{\text{s}} \leq 5.5 \times 10^5 ~\mathrm{A/m}$
\end{tabular}
&$2^{\circ}\leq\theta_{\mathrm{sk}}$$\leq 10^{\circ}$ \\
\hline
\end{tabular}
\label{tab:tab1}
\end{table}
Table~\ref{tab:tab1} summarizes the dependence of the $\theta_{\mathrm{sk}}$ on the key intrinsic material parameters-$D_{\mathrm{int}}$, $K_{\mathrm{u}}$, and $M_{\mathrm{s}}$. For stronger DMI ($D_{\mathrm{int}} \geq 3~\mathrm{mJ/m^2}$), lower anisotropy ($K_{\mathrm{u}} \leq 0.84 \times 10^6~\mathrm{J/m^3}$), and higher magnetization ($M_{\mathrm{s}} \geq 5.5 \times 10^5~\mathrm{A/m}$), $\theta_{\mathrm{sk}}$ remains below $2^{\circ}$, indicating suppressed transverse motion and enhanced dynamical stability. In contrast, a reduction in $D_{\mathrm{int}}$ and $M_{\mathrm{s}}$, together with an increase in $K_{\mathrm{u}}$, leads to larger Hall angles ($2^{\circ}$–$10^{\circ}$), reflecting stronger transverse deflection. These results establish the optimal range of material parameters for $\boldsymbol{\nabla}T(x)$-driven skyrmion dynamics, where the balance between chiral exchange and anisotropy enables precise control of skyrmion motion and trajectory.

\section{conclusion}
We have investigated thermally driven skyrmion dynamics in a Co/Pt bilayer nanoracetrack using micromagnetic simulations. Our results show that skyrmions consistently move toward the hotter side under an applied temperature gradient, with their motion governed by the interplay between magnonic and dipolar torques. In contrast to previous reports, we observe an almost vanishing skyrmion Hall effect, where the transverse velocity remains negligible compared to the longitudinal component within a specific range of material parameters, as summarized in Table~1. Systematic analysis of the damping constant, uniaxial anisotropy, interfacial DMI, and saturation magnetization reveals that while the skyrmion velocity is highly sensitive to variations in these intrinsic parameters, the skyrmion Hall angle is influenced by changes in the material parameters except for the damping constant. By extracting the skyrmion diameter as a function of these parameters-uniaxial anisotropy, DMI, and saturation magnetization, we find that the critical skyrmion size varies across material parameters, and beyond this threshold, the skyrmion dynamics changes dramatically.  In particular, when the diameter exceeds a certain value of diameter, the Hall angle is strongly suppressed while the longitudinal velocity increases. This parameter-dependent analysis establishes a robust design principle for achieving straight-line, Hall-effect-free skyrmion transport. Our findings highlight the potential of thermally driven skyrmions as low-dissipation information carriers, paving the way for energy-efficient spintronic and magnonic devices that can harness waste heat for functional operation.

\begin{acknowledgments}
We acknowledge the financial support from the Science and Engineering Research Board (SERB), Govt. of India, through grant nos. CRG/2018/004340 and SPR/2021/000762. Partial support from the Indo-French Centre for Promotion of Advanced Research (CEFIPRA) through grant no. IFC/A/6508-2/2021/489 is also acknowledged. We acknowledge Dr. Nimisha Arora for assisting in the post-analysis of magnetization images. We also thank Dr. Harish Chandra Chauhan for valuable scientific discussions. We gratefully acknowledge the High-Performance Computing (HPC) facility at IIT Delhi for providing the platform to perform micromagnetic simulations. Y.K. thanks the Ministry of Education, Government of India, for the research fellowship. H.S. acknowledges the Council of Scientific and Industrial Research (CSIR), Government of India, for financial support.
\end{acknowledgments}
\nocite{*}

\bibliography{ref}
\end{document}